\documentclass[prl,showpacs]{revtex4}
\usepackage{graphicx}
\usepackage{dcolumn}
\usepackage{bm}
\usepackage{epsfig}

\begin{document}

\title{Asymmetry of the excess finite-frequency noise}
\author{I. Safi}

\affiliation{Laboratoire de Physique des Solides, Universit\'e
Paris-Sud, 91405 Orsay, France}

\topmargin=-0.60in
\oddsidemargin=-0.125in
\textheight=9.3in
\textwidth 6.6in



\begin{abstract}
We consider finite frequency noise in a mesoscopic system with arbitrary interactions, connected to many terminals kept at finite electrochemical potentials. We show that the excess noise, obtained by subtracting the noise at zero voltage from that at finite voltage, can be asymmetric with respect to positive/negative frequencies if the system is non-linear. This explains a recent experimental observation  in Josephson junctions as well as strong asymmetry obtained in typical non-linear and strongly correlated systems described by the Luttinger liquid (LL): edge states in the fractional quantum Hall effect, quantum wires and carbon nanotubes. Another important problem where the LL model applies is that of a coherent conductor embedded in an ohmic environment.
\end{abstract}

\maketitle

The finite frequency (FF) current fluctuations have attracted a lot of interest in the mesoscopic community \cite{expe,khlus_87}. It has been possible to measure them in the quantum regime, i. e. at frequencies higher than temperature.
 Beyond the average current, they offer a powerful tool to reveal the charge and statistics of the charge carriers of a system when considered at zero-frequency. Nevertheless, they contain more rich informations at FF, such as on elementary excitations, typical energy scales, and especially on interactions, with the possibility to check the underlying model or access the correlation strength etc...
 
 It has been often stated in theoretical studies that one needs to symmetrize the current correlators, as they don't commute at different times, which leads to a FF symmetrized noise, symmetric with respect to positive/negative frequencies. Nevertheless, this has  been often explored by the scattering approach at low frequencies compared to characteristic energy scales \cite{khlus_87}. Few theoretical works went beyond this framework\cite{eugene}. It has also been studied in systems where interactions intervene at any frequency scale in the FF noise such as chiral edge states in the Fractional Quantum Hall effect (FQHE), quantum wires and carbon nanotubes, described by the Luttinger liquid (LL) model, which leads to exotic phenomena such as charge fractionalization \cite{safi}, spin-charge separation, and fractional statistics. It has been shown that LL is suited as well to describe a coherent conductor embedded into an ohmic environment\cite{safi_saleur}.  The symmetrized FF noise has been studied in the FQHE in Ref.~\cite{chamon}, and in Refs.~\cite{theo} for quantum wires and carbon nanotubes connected to metallic leads, where it was shown that, while the presence of the metallic leads obscures it in the zero-frequency noise, the charge fractionalization is still present and can be extracted from the noise at high frequencies. 

Nevertheless, recent experiments \cite{billangeon} have got access to the non-symmetrized noise \cite{creux}, and thus to the emission and the absorption components of the noise spectrum given by the noise at positive/negative frequencies. What is usually measured is the excess non-symmetrized noise, defined as the difference between the noise at finite voltage and at zero voltage, thus allowing to get rid of some undesirable effects. 
 In the framework of the scattering approach, the non-symmetrized noise is not even with respect to frequency, but the excess non-symmetrized one is, thus yields again the excess symmetrized noise. This makes it difficult to argue which noise is measured in that case, and a criteria for choosing systems with an asymmetric excess noise is required. 
 
Indeed even fewer theoretical works have dealt with the non-symmetrized FF noise beyond the framework of scattering approach \cite{hekking}. It was first investigated in strongly correlated systems such as the FQHE edge states \cite{bena}, a coherent conductor in an ohmic environment, and quantum wires and carbon nanotubes connected to charge reservoirs \cite{ines_ff_noise}. In particular, the non-symmetrized excess noise was found to be asymmetric. One could argue that such asymmetry needs Coulomb interactions to hold on, nevertheless the criteria was indeed discovered to be related to non-linearity \cite{ines_ff_noise,ines_FDT}, as we will be explained now.
 This enlightens a recent experiment measuring asymmetric FF excess noise in Josephson junctions \cite{billangeon}. 
 
For that, we need an out-of-equilibrium generalization of the Kubo formula. The latter might be thought to be restricted to linear transport. Nevertheless, it has been extended to the non-equilibrium case, looking at infinite systems subject to a uniform electric field, and relating the AC homogeneous conductivity to the retarded current-current correlation function computed in the presence of the finite DC bias \cite{gavish} with the requirement of a stationary density matrix. This misses however the effects of the non-locality, which are important in a mesoscopic context and where the electric field is not uniform for instance. A straightforward demonstration, extended in addition to any time-dependent Hamiltonian and valid generally for any finite mesoscopic non-linear system with many terminals, is presented in Ref.~\cite{ines_FDT}. Here we focus on the stationary case. We consider a general mesoscopic system connected to many terminals under DC electrochemical potentials $\mu_n=eV_n$.
Let's add a small AC modulation $v_{m}(\omega) e^{i\omega t}$ to the DC potential $V_m$ in terminal $m$. The differential of the average current at a terminal $n$ in the limit of vanishing $v_m$:
$G_{nm}(\omega)= {\delta <I_n(\omega)>}/{\delta { v}_m(\omega)}|_{{v}_m=0}$ is an element of the  AC differential conductance matrix $\bf G_V$ where ${\bf V}=(V_1,.. ,V_N)$ recalls its dependence on the DC potentials in the N terminals. One can show the generalized Kubo formula \cite{ines_FDT}:
 $
 \omega {\bf G}(\omega)={1}/{\hbar}\mathbf{D^R}(\omega)+i\mathbf{Id},
$ where $ D_{mn}^R(t)=\theta(t)\left<\left[I_{m}(t),I_n(0)\right]\right>.$ 
 As a consequence, we can show that the non-symmetrized FF noise matrix ${\bf S_{V}}(\omega)$, whose elements are given by:
\begin{eqnarray}\label{defnoise}
S_{nm}(\omega) =\int_{-\infty}^{\infty} dt e^{i\omega t}
 \left[\left\langle  I_m(0) I_n(t)
\right\rangle -<I_m><I_n>\right],
\end{eqnarray}
has its asymmetric part given by the dissipative AC differential conductance matrix: \begin{equation}\label{s-} {\bf{S}_V}(-\omega)-{\bf{S}_V}(\omega)=\hbar\omega\Re {\bf{G}_V}(\omega).
\end{equation}In particular, this is obeyed by the noise at zero voltage, i. e. at equilibrium, where the FDT yields: ${\mathbf{S}_{V=0}}(\omega)=2\hbar\omega N(\omega)\Re {\mathbf{G}_{V=0}}(\omega),$ with $N(\omega)=1/(e^{\beta\omega}+1)$.
Let's consider now the excess noise matrix, 
$
{\bf  S_V^{exc}}(\omega)={\bf  S_V}(\omega)-{\bf S_{V=0}}(\omega).
$
Taking the difference between Eq.(\ref{s-}) at finite and zero voltage vector, one gets: 
\begin{eqnarray}\label{excess_asymmetry}
{\bf { S}_V^{exc}}(-\omega)-{\bf { S}_V^{exc}}(\omega)&=&2\hbar \omega [ {\Re\mathbf{G}_V}(\omega)-{\Re\mathbf{G}_{V=0}}(\omega)].
\end{eqnarray}
This shows a necessary condition to get asymmetry in the excess noise between positive and negative frequencies: non-linearity, such that $\Re G_V$ depends on voltage! 

Let us now illustrate this in a problem relevant for interactions and environment role in mesoscopic physics, referring to \cite{bena} for details and other typical non-linear systems.
\begin{figure}[tbp]
\includegraphics[scale=0.5,clip]{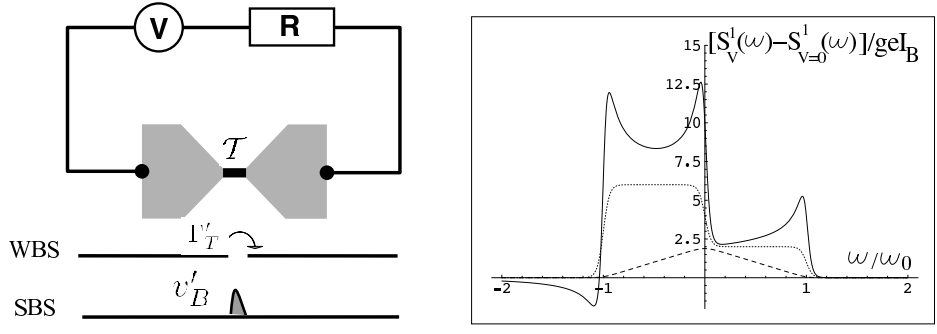}
\caption{\label{fig1}On the left: A one-channel coherent
conductor with transmission coefficient ${{\cal T}}$, in series
 with an impedance $Z(\omega)=R$ for $\omega<\omega_R=1/RC$,
 is mapped to a LL with parameter $K=1/(1+r)$ where $r=e^2 R/h$.
  The strong (weak) backscattering limit denoted by SBS (WBS) corresponds to
 the tunneling (weak backscattering) regime with a dimensionless amplitude $\Gamma'_T$
 ($v'_B$). On the right: The excess non-symmetrized FF noise  $S_V(\omega)-S_{V=0}(\omega)$ (scaled by a renormalized reflection coefficient), as a function of frequency, in units of $\omega_0=eV/\hbar$, for $R=0$ (dashed line), $R=2h/e^2$ (full line) and $R=h/e^2$ (dotted line). Here $k_B T/\hbar \omega_0=0.03$.}
 \end{figure}

A mesoscopic conductor embedded in an electrical
circuit forms a quantum system violating Ohm's laws. The
transmission/reflection processes of electrons through the
conductor excite the electromagnetic modes of the environment,
 rendering the scattering inelastic, and reducing the
current at low voltage, an effect called dynamic Coulomb
blockade (DCB)\cite{ingold_nazarov}. This picture, valid in the limit of
a weak conductance, changes in the opposite limit of a good
conductance. The description of tunneling via
discrete charge states becomes then ill defined, raising the
question of whether DCB survives or is
completely washed out by quantum fluctuations. A  
 challenging relationship between the DCB reduction of the current in a one-channel
 conductor in series with a weak impedance
 and the noise without impedance was proposed in \cite{levy_yeyati}. 
We were able to explore the case of an arbitrary
resistance $R$ in series with a coherent one channel conductor
with good transmission. At energies below $\omega_R=1/RC$ ($C$ the effective capacitance of the circuit) a one-channel
conductor embedded in its ohmic environment behaves exactly like a
point scatterer\cite{note} in a LL liquid, with an
effective LL parameter $K=1/(1+r)$ (Fig.(\ref{fig1})). 
 This allowed us to
determine in a non-perturbative way the effect of the environment on
$I-V$ curves, and to find an exact relationship between DCB and shot noise which applies for all energy ranges below $\omega_R$ and for any resistance, as well as the full counting statistics.
 Furthermore, such mapping was later confirmed using completely different approach for many channel system, but restricted to the high voltage regime \cite{zaikin_05}. It seems that our mapping could be generalized to this case, provided the resistance of the channels is incorporated in the resistance of the environment $R$, thus the LL turns out to have more universal interest.
 One can also apply the first investigation of non-symmetrized FF noise achieved in \cite{bena} in perturbation with respect to $1-\cal{T}$. Fig.(\ref{fig1}) shows strong asymmetry in the FF noise, especially when $R$ is higher than $h/e^2$ where the DCB is known to be strongest in the tunneling regime \cite{notebis}. Measuring the FF excess noise would give a crucial test for our mapping. Notice that this offers a potential and interesting alternative to differentiate the role of intrinsic interactions in quantum wires and carbon nanotubes \cite{ines_ff_noise} from the role of the environment.\\
\textit{Acknowledgments:}
The author thanks C. Bena and A. Cr\'epieux for collaborations on this subject, and E. Shukhurokov, R. Deblock and C. D. Glattli for stimulating discussions.


\begin{thebibliography}{99}
\bibitem{note} whose strength is related to the transmission coefficient $\cal{T}$.
\bibitem{notebis} The symmetrized noise is similar to that obtained in \cite{zaikin_05} for many channels.
\end{thebibliography}
\end{document}